\documentclass{article}

\usepackage[preprint]{neurips_2019}

\usepackage{CJKutf8}
\usepackage[T1]{fontenc}    
\usepackage{hyperref}       
\usepackage{url}            
\usepackage{booktabs}       
\usepackage{amsfonts}       
\usepackage{nicefrac}       
\usepackage{microtype}      
\usepackage{graphicx}
\usepackage{epstopdf}
\usepackage{setspace}
\usepackage{rotating}
\usepackage{subfig}
\usepackage{verbatim}
\usepackage{amssymb}
\usepackage{amsmath}
\usepackage{algorithm2e}
\usepackage{verbatim} 
\usepackage{epstopdf}
\usepackage{setspace}
\usepackage{amsthm}
\usepackage{rotating}
\usepackage[T1]{fontenc}
\usepackage{subfig}

\usepackage{comment}
\usepackage{hyperref}
\usepackage{tikz,xcolor,hyperref}
\usepackage{caption} 
\usepackage{hyperref}
\usepackage{setspace}
\usepackage{soul}
\definecolor{lime}{HTML}{A6CE39}
\DeclareRobustCommand{\orcidicon}{
	\begin{tikzpicture}
	\draw[lime, fill=lime] (0,0) 
	circle [radius=0.16] 
	node[white] {{\fontfamily{qag}\selectfont \tiny ID}};
	\draw[white, fill=white] (-0.0625,0.095) 
	circle [radius=0.007];
	\end{tikzpicture}
	\hspace{-2mm}
}
\foreach \x in {A, ..., Z}{%
	\expandafter\xdef\csname orcid\x\endcsname{\noexpand\href{https://orcid.org/\csname orcidauthor\x\endcsname}{\noexpand\orcidicon}}
}

\usepackage[compact]{titlesec} 
\title{Genotyping coronavirus SARS-CoV-2: methods and implications}
\author{%
   Changchuan Yin \thanks{Correspondence author, cyin1@uic.edu \orcidA{}}\\\\
   Department of Mathematics, Statistics, and Computer Science \\
   University of Illinois at Chicago \\
   Chicago, IL 60607 \\
   USA\\
}
\begin{document}
\maketitle
\begin{abstract}
The emerging global infectious COVID-19 coronavirus disease by novel Severe Acute Respiratory Syndrome Coronavirus 2 (SARS-CoV-2) presents critical threats to global public health and the economy since it was identified in late December 2019 in China. The virus has gone through various pathways of evolution. For understanding the evolution and transmission of SARS-CoV-2, genotyping of virus isolates is of great importance. We present an accurate method for effectively genotyping SARS-CoV-2 viruses using complete genomes. The method employs the multiple sequence alignments of the genome isolates with the SARS-CoV-2 reference genome. The SNP genotypes are then measured by Jaccard distances to track the relationship of virus isolates. The genotyping analysis of SARS-CoV-2 isolates from the globe reveals that specific multiple mutations are the predominated mutation type during the current epidemic. Our method serves a promising tool for monitoring and tracking the epidemic of pathogenic viruses in their gradual and local genetic variations. The genotyping analysis shows that the genes encoding the S proteins and RNA polymerase, RNA primase, and nucleoprotein, undergo frequent mutations. These mutations are critical for vaccine development in disease control. 
\end{abstract}
\section{Highlights}
\begin{itemize}
	\item We genotyped 558 SARS-CoV-2 isolates from the globe as of March 23, 2020.
    \item Frequent mutations in SARS-CoV-2 genomes are in the genes encoding the S protein and RNA polymerase, RNA primase, and nucleoprotein.
    \item We established a method for monitoring and tracing SARS-CoV-2 mutations. 
\end{itemize}
\section{Introduction}
\label{Introduction}
The novel coronavirus in humans, first discovered in Wuhan, China, in December 2019, was initially named as 2019-nCoV and then designated as SARS-CoV-2 due to its taxonomic and genomic relationships with the species Severe acute respiratory syndrome-related coronavirus \citep{gorbalenya2020species}. The present outbreak of the coronavirus-associated acute respiratory disease is named coronavirus disease 19 (COVID-19) by WHO. Since the epidemic of COVID-19, more than 332, 930 people from 147 countries and territories have been confirmed sicked and more than 14, 510 have died from the rapidly-spreading SARS-CoV-2 virus as of March 23, 2020 \citep{who_2020}. 

Coronaviruses (CoVs) are a family of enveloped positive-strand RNA viruses infecting vertebrates, named for the crown-like spikes on their surface. Coronavirus (CoV) belongs to the family\textit{ Coronaviridae} and the order \textit{Nidovirales}. Coronavirus is widely spread in humans, other mammals, and birds, and can cause diseases such as the respiratory, intestinal, liver, and nervous systems. Human coronaviruses (HCoVs) were first identified in the mid-1960s. Seven common HCovs are CoV-229E (alpha coronavirus), CoV-NL63 (alpha coronavirus), CoV-OC43 (beta coronavirus), CoV-HKU1 (beta coronavirus), Severe acute respiratory syndrome coronavirus (SARS-CoV), Middle East respiratory syndrome coronavirus (MERS-CoV), and current SARS-CoV-2. CoV-229E and CoV-OC43 are the cause of the common cold in adults during the mid-1960s. Disease manifestations associated with CoV-HKU1 and CoV-NL63 include the common cold and chronic pneumonia. Coronavirus-HKU1 has been predominantly reported in children in the United States but less common among adults. Three highly pathogenic coronaviruses, SARS-CoV, MERS-CoV, and SARS-CoV-2, which emerged in 2002, 2012, and 2019, respectively, have caused severe respiratory disease and thousands of deaths worldwide \citep{chen2020pathogenicity}.

SARS-CoV-2 coronavirus harbors a linear single-stranded positive RNA genome. The coronavirus SARS-CoV-2 genome consists of a leader sequence, ORF1ab encoding proteins for RNA replication, and genes for non-structural proteins (nps) and structural proteins. The genomic leader sequence of about 265 bp is the unique characteristic in coronavirus replication and plays critical roles in the gene expression of coronavirus during its discontinuous sub-genomic replication \citep{li2005sirna}. ORF1ab encodes replicase polyproteins required for viral RNA replication and transcription \citep{chen2020emerging}. Expression of the C-proximal portion of ORF1ab requires (–1) ribosomal frame-shifting. The first non-structural protein (nsp) encoded by ORF1ab is Papain-like proteinase (PL proteinase, nps3). Nsp3 is an essential and largest component of the replication and transcription complex. The PL proteinase in nsp3 cleaves nsps 1-3 and blocks host innate immune response, promoting cytokine expression \citep{serrano2009nuclear,lei2018nsp3}. Nsp4 encoded in ORF1ab is responsible for forming double-membrane vesicle (DMV). The other nsp are 3CLPro protease (3-chymotrypsin-like proteinase, 3CLpro) and nsp6. 3CLPro protease is essential for RNA replication. The 3CLPro proteinase is accountable for processing the C-terminus of nsp4 through nsp16 in all coronaviruses \citep{anand2003coronavirus}. Therefore, conserved structure and catalytic sites of 3CLpro may serve as attractive targets for antiviral drugs \citep{kim2012broad}. Together, nsp3, nsp4, and nsp6 can induce DMV \citep{angelini2013severe}.

SARS-coronavirus RNA replication is unique, involving two RNA-dependent RNA polymerases (RNA pol). The first RNA polymerase is primer-dependent non-structural protein 12 (nsp12), and the second RNA polymerase is nsp8. In contrast to nsp12, nsp8 has the primase capacity for \textit{de novo} replication initiation without primers \citep{te2012sars}. Nsp7 and nsp8 are important in the replication and transcription of SARS-CoV-2. The SARS-coronavirus nsp7 and nsp8 complex is a multimeric RNA polymerase for both \textit{de novo} initiation and primer extension \citep{prentice2004identification, te2012sars}. Nsp8 also interacts with ORF6 accessory protein. Nsp9 replicase protein of SARS-coronavirus binds RNA and interacts with nsp8 for its functions \citep{sutton2004nsp9}. 

Furthermore, the SARS-CoV-2 genome encodes four structural proteins. The structural proteins possess much higher immunogenicity for T cell responses than the non-structural proteins \citep{li2008t}. The structural proteins are involved in various viral processes, including virus particle formation. The structural proteins include spike (S), envelope (E), membrane protein (M), and nucleoprotein (N), which are common to all coronaviruses \citep{marra2003genome,ruan2003comparative}. The spike S protein is a glycoprotein, which has two domains S1 and S2. Spike protein S1 attaches the virion to the cell membrane by interacting with host receptor ACE2, initiating the infection \citep{wong2004193}. After the internalization of the virus into the endosomes of the host cells, the S glycoprotein is induced by conformation changes. The S protein is then cleaved by cathepsin CTSL, and unmasked the fusion peptide of S2, therefore, activating membranes fusion within endosomes. Spike protein domain S2 mediates fusion of the virion and cellular membranes by acting as a class I viral fusion protein. Especially, the spike glycoprotein of coronavirus SARS-CoV-2 contains a furin-like cleavage site \citep{coutard2020spike}. The furin recognition site is important for being recognized by pyrolysis and therefore, contributing to the zoonotic infection of the virus. The envelope (E) protein interacts with membrane protein M in the budding compartment of the host cell. The M protein holds dominant cellular immunogenicity \citep{liu2010membrane}. Nucleoprotein (ORF9a) packages the positive-strand viral RNA genome into a helical ribonucleocapsid (RNP) during virion assembly through its interactions with the viral genome and membrane protein M \citep{he2004characterization}. Nucleoprotein plays an important role in enhancing the efficiency of subgenomic viral RNA transcription as well as viral replication.

The increasing epidemiological and clinical evidence implicates that the SARS-CoV-2 has stronger transmission power than SARS-CoV and lower pathogenicity \citep{guan2020clinical}. However, the mechanism of high transmission of SARS-CoV-2 is unclear. DNA sequence comparisons using single nucleotide polymorphisms (SNPs) are often used for evolutionary studies and can be especially beneficial in recognizing the mutated coronavirus genomes, where high mutations can occur due to an error-prone RNA-dependent RNA polymerase in genome replication.

To understand the virus evolution of SARS-CoV-2 from the genome mutation context, we establish the SNP genotyping method and investigate the genotype changes during the transmission of SARS-CoV-2. Our results show that the genotypes of the virus are not uniformly distributed among the complete genomes of SARS-CoV-2. This genotyping study discovers a few highly frequent mutations in the SARS-CoV-2 genomes. The highly frequent SNP mutations might be associated with the changes in transmissibility and virulence of the virus. The mutations are located in the S protein, RNA polymerase, RNA primase, and nucleoprotein, which are fundamental proteins for vaccine efficacy. Therefore, the high-frequency SNP mutations are important factors when developing vaccines for preventing the infection of SARS-CoV-2 coronavirus. 

\section{Methods and algorithms}
\subsection{Multiple sequence alignments (MSA)}
Total 558 complete genome sequences of the SARS-CoV-2 strains from the infected individuals are retrieved from the GISAID database \citep{shu2017gisaid} as of March 23, 2020. Only the complete genomes of high-coverage are included in the dataset. The countries and territories, which are infected by SARS-CoV-2 and share the complete genomes of SARS-COV-2, are Australia (AU), Belgium (BE), Brazil (BR), Canada (CA), Chile (CL), China (CN), Czech Republic (CZ), Denmark (DK), England (UK), Finland (FI), France (FR), Georgia (GE), Germany (DE), Hong Kong (HK), Hungary (HU), India (IN), Ireland (IE), Italy (IT), Japan (JP), Korea (KR), Kuwait (KW), Mexico (MX), Netherlands (NL), New Zealand (NZ), Scotland (UK), Singapore (SG), Switzerland CH), Sweden(SE), Taiwan (TW), Thailand (TH), United Kingdom (UK), Unites States (US), and Vietnam (VN). The complete genome sequences are aligned with the reference genome of SARS-CoV-2 by MSA tool Clustal Omega using the default parameters \citep{sievers2014clustal}. The aligned genomes are then re-positioned according to the reference SARS-CoV-2 genome (GenBank access number: NC\_045512.2). 

\subsection{SNP genotyping}
The SNP mutations including nucleotide changes and the corresponding positions in a genome are called an SNP profile. The SNP profiles of SARS-CoV-2 isolates are retrieved and parsed from the aligned genomes according to the reference genome SARS-CoV-2. The SNP profile of the complete genome of a virus can be considered as the genotype of the virus.  

\subsection{Jaccard distance of the SNP variants}
The Jaccard similarity coefficient $J(A, B)$ of two sets $A$ and $B$ is defined as the intersection size of the two sets divided by the union size of two sets (Equation (1)) \citep{levandowsky1971distance}.
\begin{equation}
J(A,B) = \frac{{\left| {A \cap B} \right|}}
{{\left| {A \cup B} \right|}} = \frac{{\left| {A \cap B} \right|}}
{{\left| A \right| + \left| B \right| - \left| {A \cap B} \right|}}
\end{equation}
The Jaccard distance is a metric on the collection of finite sets. The Jaccard distance $d_J (A, B)$ of two sets $A$ and $B$ is scored by the difference between 100\% and the Jaccard similarity coefficient (Equation (2)). 
\begin{equation}
	d_J (A, B) = 1 - J(A,B) = \frac{{\left| {A \cup B} \right| - \left| {A \cap B} \right|}}
	{{\left| {A \cup B} \right|}}
\end{equation}
The Jaccard distance measure of SNP variants takes account of the ordering of SNP mutations. Therefore, the genetic distance of two genomes corresponds to the Jaccard distance of their SNP variants. The Jaccard distance of SNP variants was adopted in the phylogenetic analysis of human or bacterial genomes \citep{comas2009genotyping, yu2017novel, yin2019whole}. In this study, we use the Jaccard distance of the SNP mutations of virus genomes to measure the dissimilarity of virus isolates.  
\subsection{Transmission analysis of virus isolates by SNP genotyping}
Because a mutation is rarely reversed, more SNPs in a virus occur along time. Let $A$ and $B$ represent two SNP sets of the virus, if $A$ is the subset of $B$, i.e., $(A \in B, A \ne B)$, then B can be considered as one of A's descendants $A$, and $A$ can be considered as the ancestor of $B$. To this end, we propose the directed Jaccard distance $D_J (A, B)$ of two SNP sets $A$ and $B$ as the measure of mutual relationship (Equation (3)). Obviously, if $B$ is a descendant of $A$, then $D_J (A, B)$ is positive; otherwise, if $A$ is a descendant of $B$, $D_J (A, B)$ is negative. In all the descendants of an SNP $A$, the closest descendant is the one having the minimum $D_J (A, B)$ of the $A$ descendant sets.
\begin{equation}
D_J (A,B) = \operatorname{sgn} (1 - J(A,B)) = \left\{ \begin{gathered}
\frac{{\left| {A \cup B} \right| - \left| {A \cap B} \right|}}
{{\left| {A \cup B} \right|}},{\kern 1pt} {\kern 1pt} {\text{if}}{\kern 1pt} A \cap B \cong A \hfill \\
\frac{{\left| {A \cap B} \right| - \left| {A \cup B} \right|}}
{{\left| {A \cup B} \right|}},{\kern 1pt} {\text{if}}{\kern 1pt} A \cap B \cong B \hfill \\ 
\end{gathered}  \right.
\end{equation}
For two SNP sets $A$ and $B$, if $A \cap B \ne \emptyset, A \not\subset B$ and $B \not\subset A$, then the two viruses are relatives, sharing common SNP mutations. If two SNP sets are neither descendant-ancestor nor relatives, the corresponding two viruses are isolated mutants. Hence, the relevance of virus isolates can be identified from the directed Jaccard measure on the SNP genotypes.

Though the source of SARS-CoV-2 varies, we still consider the virus samples were randomly collected for sequencing. If a virus strain among all sequenced viruses has many descendants in the genome set, we infer that this strain is conferred with high transmissibility. Therefore, the SNP mutations in this strain are critical for increased transmissibility.
 
We calculate the directed Jaccard distances of the SNP mutations to identify the relationships of virus strains, therefore, we may determine the virus transmission pattern. The pipeline for SNP genotyping and analysis is described in Algorithm 1. 

\begin{algorithm} 
	\SetAlgoLined
	\KwIn{The complete genomes of SARS-CoV-2 strains}
	\KwOut{SNP genotypes of SARS-CoV-2 strains}
	\textbf{Step:}
	\begin{enumerate}
		\item Divide the complete genomes of SARS-CoV-2 strains into subsets based on the originating territories.
		\item Add the reference genome of SARS-CoV-2 to each subset of the complete genomes. 
		\item Perform multiple sequence alignments for each subset genomes using Clustal Omega.
		\item Convert the alignment files to SNP profiles using the reference genome of SARS-CoV-2.
		\item Merge the SNP profiles of all virus genomes.
		\item Calculate the pairwise directed Jaccard distances of all the SNPs profiles.
		\item Analyze the descendants, ancestors, and relative relationships of each SNP genotype from the Jaccard distances.
	\end{enumerate}

	\caption{SNP genotyping analysis of SARS-CoV-2.}
\end{algorithm}

\subsection{Data and computer programs}
The genomic analytics is performed using computer programs in Python and Biopython libraries \citep{cock2009biopython}. The computer programs and the updated SNP profiles of SARS-CoV-2 isolates are available upon requests. 

\section{Results}
\subsection{Genotyping SARS-CoV-2 coronavirus isolates from the globe}
We retrieve the SNP genotypes of 442 SARS-CoV-2 strains in GISAID database from the globe. To investigate the SNP distributions among all the virus isolates, we plot the SNP profiles of all the virus isolates from the globe and compare the frequency of each SNP mutation in the virus sets. The results show large mutation diversity in these virus isolates. 

From the mutation frequency analysis, the mutations are due to the fact that RNA-dependent RNA polymerase (RdRp) of RNA viruses lacks proofreading, however, the mutations are not equally distributed. The SNP mutations can be single mutation and multiple mutations at a few fixed positions. The impacts and roles that these SNP mutations have on the pathogenicity and transmission ability of SARS-CoV remain to be determined by biochemical experiments. These divers mutations might impact both transmissibility and pathogenicity of SARS-CoV-2.
\begin{figure}[tbp]
	\centering
	\subfloat[]{\includegraphics[width=4.25in]{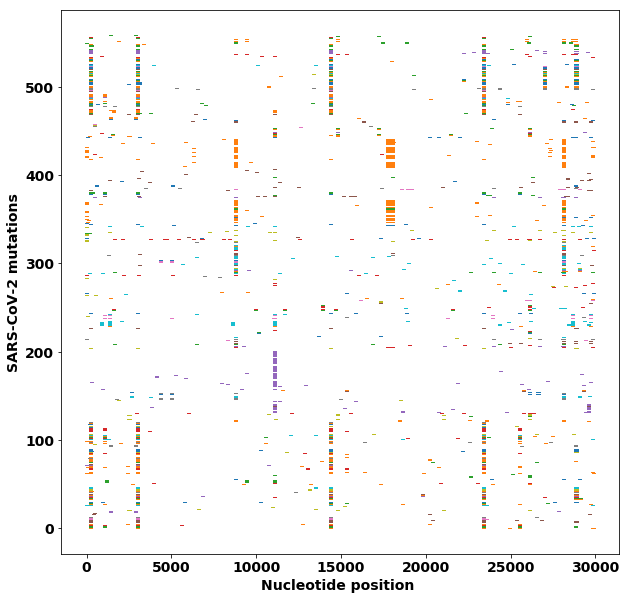}}\quad
	\subfloat[]{\includegraphics[width=4.25in]{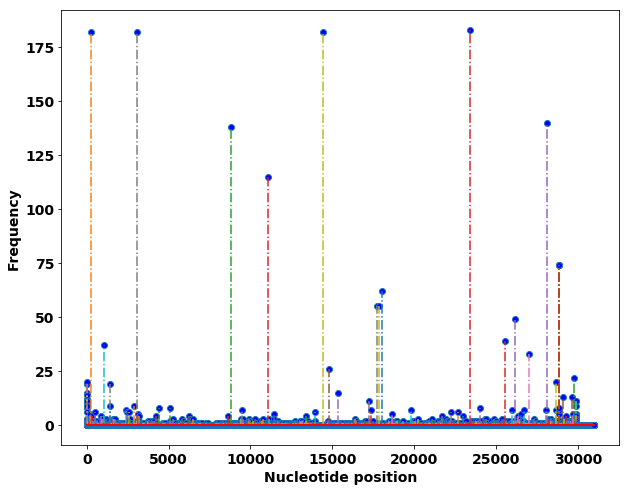}}\quad
	\caption{Distribution of SNP mutations of SARS-CoV-2 isolates from the globe. (a) The SNP profiles of mutations in 442 SARS-CoV-2 isolates. (b) Frequencies of the single SNP mutations on the genome. The nucleotide positions are on the reference genome of SARS-CoV-2.}
	\label{fig:sub1}
\end{figure}

The first common SNP mutation in the SARS-CoV-2 genome is in the leader sequence (241C>T), an important genomic site for discontinuous sub-genomic replication. The leader sequence mutation 241C>T is co-evolved with three important mutations, 3037C>T, 14408C>T, and 23403A>G, which result in amino acid mutations in nsp3 (synonymous mutation), RNA primase (P323L), and spike glycoprotein (S protein, D614G), respectively. These three co-mutations (241C>T, 14408C>T, and 23403A>G) are in critical proteins for RNA replication (241C>T, 14408C>T) and the S protein (23403A>G) for binding to ACE2 receptor. We observe that these four co-mutations are prevalent in the virus isolates from Europe, where infections COVID-19 by SARS-CoV-2 are generally more severe than other geographical regions. Combined, these four co-mutations probably can confer increased transmissibility of the virus.  

SARS-coronavirus RNA replication is unique, involving two RNA-dependent RNA polymerases (RdRp). The first RNA polymerase is primer-dependent non-structural protein 12 (nsp12), whereas the second RNA polymerase is nsp8. Nsp8 has the primase capacity for \textit{de novo} initiation RNA replication without primers \citep{te2012sars}. The most abundant SNP mutation in SARS-CoV-2 isolates is (28144T>C) in nsp8 protein, in which amino acid leucine (L) is mutated to serine (S). Our result is consistent with a previous study on 103 SARS-CoV-2 genomes in which SARS-CoV-2 virus is classified as S and L types by the two co-mutations (8782C>T and 28144T>C) \citep{zhang2020origin}. 

The third abundant SNP mutation is (26144G>T) in nonstructural protein 3 (nsp3: G251V). The protein nsp3 works with nsp4 and nsp6 to induce double-membrane vesicles (DMV),  membrane complex that acts as a platform for RNA replication and assembly \citep{angelini2013severe}. 

The significant SNP mutation (23403A>G) is located in the gene encoding spike glycoprotein (S protein: D614G). The S protein in the SARS-CoV-2 virus is an important determinant of the host range and pathogenicity. The S protein attaches the virion to the cell membrane by binding with the host ACE2 receptor \citep{xiao2003sars}. The mutation D614G is located in the putative S1–S2 junction region near the furin recognition site (R667) for the cleavage of S protein when the viron enters or exists cells \citep{follis2006furin}. However, the actual functional impact of this high-frequency SNP mutation (23403A>G) in the S protein (D614G) is unclear. The affinity strength of the mutation S protein (D614G) with the ACE2 receptor shall be further determined by biochemical experiments. 

Especially, the SNP analytics result also shows that the primer independent RNA primase (nsp8) contains more mutations than any other proteins (28144T>C, 28881G>A, 28881G>A, 28882G>A, and 28883G>C). The RNA polymerase and primase mutations may confer resistance to mutagenic nucleotide analogs via increased fidelity. The previous study indicated that a single mutation in RNA polymerase can improve the replication fidelity in RNA virus \citep{pfeiffer2003single}. If a mutation is lethal or reduces the transmission ability, the mutations may not be carried on or get deceased. The SNP profiles demonstrate that the mutations in the envelope glycoprotein and RNA polymerases predominate. Only the mutations in the S protein that have strongly binding to cell ACE2 receptors while escape from immune system response can have chances to survive. Therefore, these critical mutations are the results of natural selection in virus evolution.

In the SARS-CoV-2 strains found in the US, the nucleocapsid  (N) protein gene has three mutations (28881G>A, 28882G>A, and 28883G>C), The N protein of SARS-CoV is responsible for the formation of the helical nucleocapsid during virion assembly. The N protein may cause an immune response and has potential value in vaccine development \citep{zhao2005immune}. These mutations shall be considered when developing a vaccine using the N protein. 

\begin{table}[ht]
\caption{High-frequency single SNP genotypes in SARS-CoV-2.}
	\centering 
	\begin{tabular}{lllllr}
		\hline\hline
		\noalign{\vskip 0.05in}   
		SNP mutation & protein mutation & frequency &  \\
		\hline
		\noalign{\vskip 0.05in}   
		241C>T & leader sequence & 178 &  \\ 
		3037C>T & synonymous mutation (nsp3, F105F) & 182 &  \\ 
		8782C>T & synonymous mutation (nsp4, S75S) & 138 &  \\ 
		11083G>T & nsp6, L37F & 115 &  \\ 
		14408C>T & RNA pol (nsp12, P323L) & 182&  \\ 
		17747C>T & helicase, P504L & 55 &  \\
		17858A>G & helicase, Y541C & 55 &  \\
		18060C>T & synonymous mutation (3'-to-5'exonuclease, L6L) & 62 &  \\
		23403A>G & spike glycoprotein (S protein), D614G & 183&  \\ 
		26144G>T & ORF3a, G251V & 49 &  \\ 
		27046C>T & membrane glycoprotein, T175M & 33 &  \\ 
		28144T>C & RNA primase (nsp8, L84S) & 140 &  \\ 
		28881G>A & nucleocapsid phosphoprotein (R203K) & 74 &  \\ 
		28882G>A & nucleocapsid phosphoprotein (R202R) & 74 &  \\ 
		28883G>C & nucleocapsid phosphoprotein (G204R) & 74 &  \\ 
		\hline\hline
		\noalign{\vskip 0.05in}   
		\multicolumn{3}{l}{%
		\begin{minipage}{5.0in}%
		\small Note: The SNP mutation positions are on the reference genome. Nucleotide T represents nucleotide U in SARS-CoV-2 RNA virus genome. The frequencies of mutations are computed from total 558 SARS-CoV-2 strains. 
		\end{minipage}%
		}\\
	\end{tabular}
	\label{table:nonlin} 	
\end{table}

\begin{table}[ht]
	\caption{Co-mutations with high descendants in SARS-CoV-2.}
	\centering 
	\begin{tabular}{l*{2}{l}r}
		\hline\hline
		\noalign{\vskip 0.05in}   
		SNP co-mutations & proteins & descendants \\
		\hline
		\noalign{\vskip 0.05in}   
		8782C>T, 28144T>C, 18060C>T>C & RNA pol (nsp8) & 54\\
		241C>T, 3037C>T, 23403A>G, 28144T>C, & S protein, RNA pol (nsp8) & 82 \\
		241C>T, 3037C>T, 14408C>T, 23403A>G & RNA primase (nsp12), S protein & 81 \\ 
    	\hline\hline
		\noalign{\vskip 0.05in}   
		\multicolumn{3}{l}{%
			\begin{minipage}{5.0in}%
				\small Note: The SNP mutation positions are on the reference genome. Nucleotide T represents nucleotide U in SARS-CoV-2 RNA genome. The frequencies of mutations are computed from total 558 SARS-CoV-2 strains. 
			\end{minipage}%
		}\\
	\end{tabular}
	\label{table:nonlin} 
  \end{table}

\subsection{Evolution of SARS-CoV-2 coronavirus by genotyping}
To spread, a pathogen virus must multiply within the host to ensure transmission, while simultaneously avoiding host morbidity or death. Therefore, during the evolution of a virus, the transmissibility of the virus is usually increased, whereas the pathogenicity becomes reduced \citep{alizon2009virulence}. From the SNP profiles of SARS-CoV-2 strain, high-frequency mutations predominate in the virus isolations, therefore, these high-frequency mutations probably contribute to increased transmissibility. In addition, these high-frequency mutations are associated with different critical proteins. We analyze and trace the SNP profiles from 442 SARS-CoV-2 strains which have at least 10 descendants. The result suggests a number of high-frequency mutations that are associated with different critical proteins. The results show that the SNP distribution is not random but is predominated at some positions and then have more descendants. These high-frequency mutations may confer a high transmissibility of the virus (Table 2). If we exclude the leader sequence mutation and the synonymous mutations (3037C>T, 8782C>T, 18060C>T), we classify the SNP mutations into four major groups based on the impacted proteins (Fig. 2.). (1) single mutation in nsp6 (11083G>T) (Fig.2(a)), (2) single mutation in ORF3a (26144G>T) (Fig.2(b)), (3) single mutation in RNA polymerase (nsp8) (8782C>T, 28144T>C) (Fig.2(c)), and (4) double mutations in S-protein and RNA polymerase: (241C>T, 3037C>T, 14408C>T, 23403A>G ) (Fig.2(d)). These strains in one group are derived from the same ancestor stain in that group according to their SNP profiles.

\begin{figure}[tbp]
	\centering
	\subfloat[]{\includegraphics[width=2.5in]{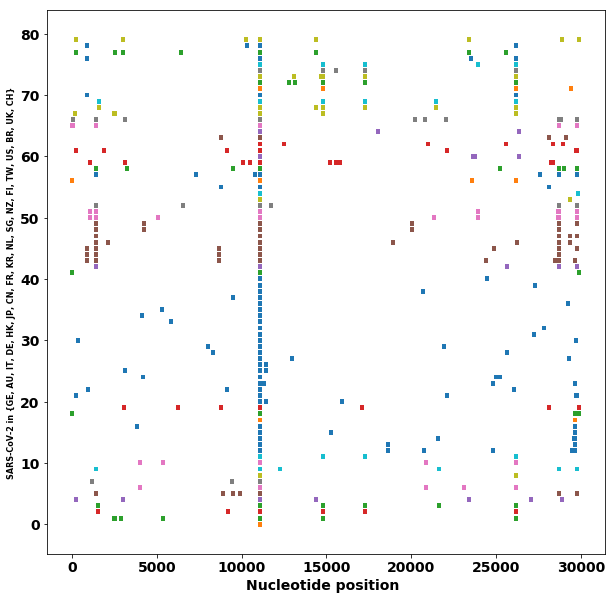}}
	\subfloat[]{\includegraphics[width=2.5in]{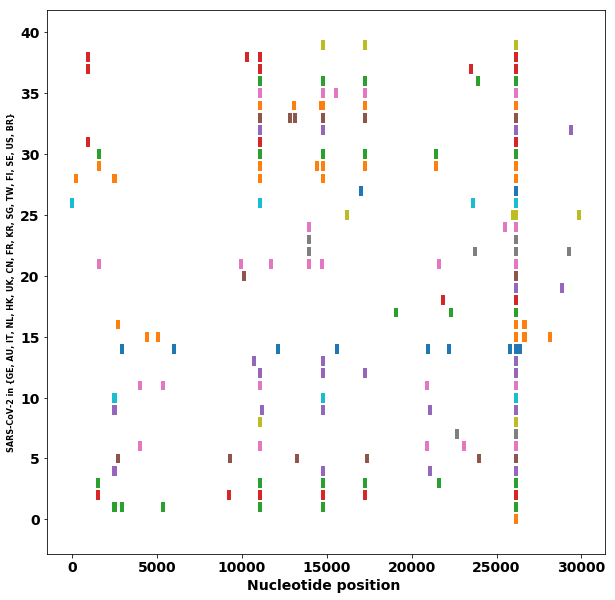}}\quad
	\subfloat[]{\includegraphics[width=2.5in]{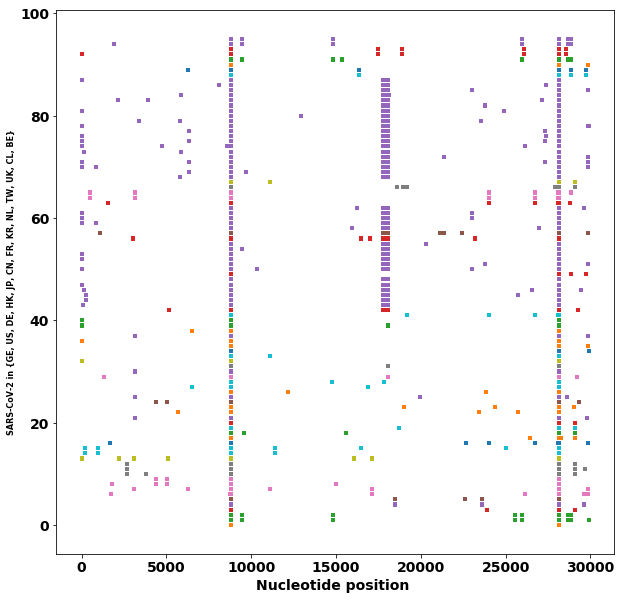}}
	\subfloat[]{\includegraphics[width=2.5in]{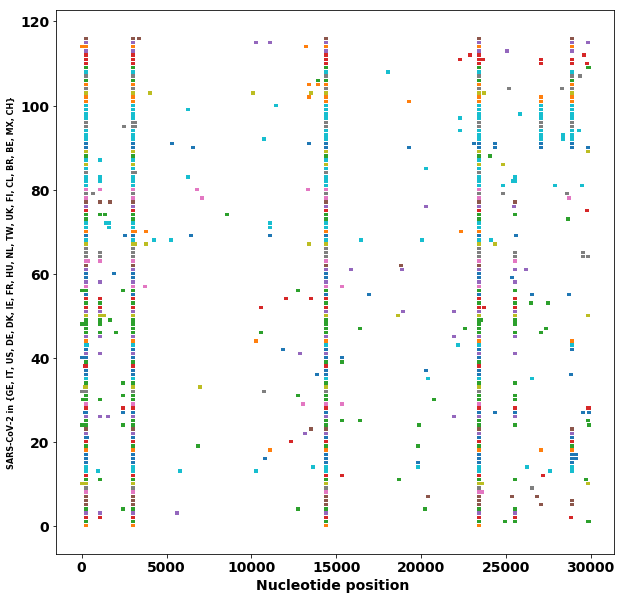}}\quad
	\caption{The SNP profiles of four major genotypes. (a) Genotype I (11083G>T), (b) Genotype II: (26144G>T), (c) Genotype III (8782C>T, 28144T>C), (d) Genotype IV (241C>T, 3037C>T, 14408C>T, 23403A>G). The strains in a genotype group originate from the same ancestor. The strains from the same region are marked in the same color.}
	\label{fig:sub1}
\end{figure}

The result shows that most SNP mutations in SARS-CoV-2 isolates in China and some from Europe and USA are located at two positions (8782C>T, 28144T->C) (Fig.2(c)). Later on this strain was mutated at new position (8782C>T, 28144T>C, 18060C>T). These mutations are from the early phase of the strain. 
   
The important and prevalent co-mutations (241C>T, 3037C>T, 23403A>G) occurred mostly in  SARS-CoV-2 isolates in Europe countries. This strain then has additional extended mutations at positions (241C>T, 3037C>T, 14408C>T, 23403A>G) (Fig.2(d)). The impacted critical proteins are NA pol (nsp8), RNA primase (nsp12), and the S protein. Most of the strains are found in Europe countries (Fig.2(d)). Italy is being heavily infected by SARS-CoV-2 with 59, 138 confirmed cases and 5, 476 deaths as of March 23, 2020 \citep{who_2020}. These critical mutations probably may be correlated with the severe infections in Europe.

From the SNP profiles of the viruses across the globe from a different time, we may estimate that one mutation can occur in one generation. For example, in USA (IL) two consecutive infection cases (US|IL1|EPI\_ISL\_404253|2020-01-21,US|IL2|EPI\_ISL\_410045|2020-01-28), the virus increased one mutation (28854C>Y) between two same community members. Over the length of its 30kb genome, SARS-CoV-2 may accumulate mutations ranging from single mutation to 14 mutations (NL|EPI\_ISL\_413591|2020-03-02), as seen from December 2019 to March 23, 2020. Therefore, we may estimate that the transmission of SARS-CoV-2 has reached 14 generations since its first infection to humans in December 2019.

Besides the SNPs mutations, we also observed a few deletion or insertion mutations in SARS-CoV-2 isolates. The deletion-insertion mutations do not happen often, however, whether these deletion and insertion mutations can spread is unknown from the limited genome data. 

\section{Discussions}
Our study has a few notable limitations due to the nature of the genome data. Because the sample collection dates may not reflect the actual infection date so the transmission path analysis is only approximate. Caution should be exercised on the genotyping analytics because some countries have not sequenced enough virus samples, the frequencies of the genotype groups may be unbalanced due to the unavailability of complete genomes in some countries and regions. Whether any of these common SNP mutations will result in biological and clinical differences remains to be determined. 

In this study, we use the complete genomes of SARS-CoV-2 for SNP genotype calling. However, in an emergency time, the complete genomes may not be available for SNP genotyping. In this case, the SNP variant calling process may directly use the raw NGS reads \citep{yin2019whole}. The SNP variants then can be obtained by mapping the NGS reads to the reference genome by BWA alignments \citep{li2013aligning}, followed by GATK variant calling \citep{mckenna2010genome}. 

\section{Conclusion}
The SARS-CoV-2 epidemic has caused a substantial health emergency and economic stress in the world. Therefore, understanding the nature of this virus and deriving methods to monitor the spread of virus in the epidemic are critical in disease control. Our results show several molecular facets of the SARS-CoV-2 pertinent to this epidemic. The discovery of genotypes linked to geographic and temporal clusters of infectious suggests that genome SNP signatures can be used to track and monitor the epidemic.

Rapid detection of different genotypes of SARS-CoV-2 are important for an efficient response to the COVID-19 outbreak Discriminating and relating viral isolates can be useful in genetic epidemiology. Determining the origin and monitoring the transmission pattern of the pathogenic agents are critical to controlling the outbreak. In this work, the SNP genotyping of SARS-CoV-2 was developed by adapting fast MSA of the complete genomes of SARS-CoV-2 and SNP analytics using the directed Jaccard distance of the SNP profiles. The genotyping analysis provides insights on the frequent mutations that confer fast transmissibility of the virus. The major mutations are in the critical proteins, including the S protein, RNA polymerase, RNA primase, and nucleoprotein. Therefore, these high-frequency SNP mutation sites must be considered when designing a vaccine for preventing the infection of SARS-CoV-2. 

\section{Abbreviations}
\begin{itemize}
	  \item COVID-19: coronavirus disease 2019 
	  \item DMV: double-membrane vesicle
	  \item GATK: the genome analysis toolkit 
	  \item MSA: multiple sequence alignment 
	  \item NGS: next generation sequencing
	  \item SARS: severe acute respiratory syndrome
	  \item SARS-CoV-2: severe acute respiratory syndrome coronavirus 2
	  \item SNP: single nucleotide polymorphisms
	  \item WHO: the world health organization
\end{itemize}

\section{Acknowledgments}
We sincerely appreciate the researchers worldwide who sequenced and shared the complete genome data of SARS-CoV-2 and other coronaviruses from
GISAID (https://www.gisaid.org/). This research is dependent on these precious data. 
\clearpage
\bibliographystyle{elsarticle-harv}
\bibliography{../References/myRefs}

\begin{thebibliography}{36}
\expandafter\ifx\csname natexlab\endcsname\relax\def\natexlab#1{#1}\fi
\providecommand{\url}[1]{\texttt{#1}}
\providecommand{\href}[2]{#2}
\providecommand{\path}[1]{#1}
\providecommand{\DOIprefix}{doi:}
\providecommand{\ArXivprefix}{arXiv:}
\providecommand{\URLprefix}{URL: }
\providecommand{\Pubmedprefix}{pmid:}
\providecommand{\doi}[1]{\href{http://dx.doi.org/#1}{\path{#1}}}
\providecommand{\Pubmed}[1]{\href{pmid:#1}{\path{#1}}}
\providecommand{\bibinfo}[2]{#2}
\ifx\xfnm\relax \def\xfnm[#1]{\unskip,\space#1}\fi
\bibitem[{Alizon et~al.(2009)Alizon, Hurford, Mideo and
  Van~Baalen}]{alizon2009virulence}
\bibinfo{author}{Alizon, S.}, \bibinfo{author}{Hurford, A.},
  \bibinfo{author}{Mideo, N.}, \bibinfo{author}{Van~Baalen, M.},
  \bibinfo{year}{2009}.
\newblock \bibinfo{title}{Virulence evolution and the trade-off hypothesis:
  history, current state of affairs and the future}.
\newblock \bibinfo{journal}{Journal of Evolutionary Biology}
  \bibinfo{volume}{22}, \bibinfo{pages}{245--259}.
\bibitem[{Anand et~al.(2003)Anand, Ziebuhr, Wadhwani, Mesters and
  Hilgenfeld}]{anand2003coronavirus}
\bibinfo{author}{Anand, K.}, \bibinfo{author}{Ziebuhr, J.},
  \bibinfo{author}{Wadhwani, P.}, \bibinfo{author}{Mesters, J.R.},
  \bibinfo{author}{Hilgenfeld, R.}, \bibinfo{year}{2003}.
\newblock \bibinfo{title}{Coronavirus main proteinase {(3CLpro)} structure:
  basis for design of anti-{SARS} drugs}.
\newblock \bibinfo{journal}{Science} \bibinfo{volume}{300},
  \bibinfo{pages}{1763--1767}.
\bibitem[{Angelini et~al.(2013)Angelini, Akhlaghpour, Neuman and
  Buchmeier}]{angelini2013severe}
\bibinfo{author}{Angelini, M.M.}, \bibinfo{author}{Akhlaghpour, M.},
  \bibinfo{author}{Neuman, B.W.}, \bibinfo{author}{Buchmeier, M.J.},
  \bibinfo{year}{2013}.
\newblock \bibinfo{title}{Severe acute respiratory syndrome coronavirus
  nonstructural proteins 3, 4, and 6 induce double-membrane vesicles}.
\newblock \bibinfo{journal}{MBio} \bibinfo{volume}{4},
  \bibinfo{pages}{e00524--13}.
\bibitem[{Chen(2020)}]{chen2020pathogenicity}
\bibinfo{author}{Chen, J.}, \bibinfo{year}{2020}.
\newblock \bibinfo{title}{Pathogenicity and transmissibility of {2019-nCoV} —
  a quick overview and comparison with other emerging viruses}.
\newblock \bibinfo{journal}{Microbes and Infection} \bibinfo{volume}{22},
  \bibinfo{pages}{69--71}.
\bibitem[{Chen et~al.(2020)Chen, Liu and Guo}]{chen2020emerging}
\bibinfo{author}{Chen, Y.}, \bibinfo{author}{Liu, Q.}, \bibinfo{author}{Guo,
  D.}, \bibinfo{year}{2020}.
\newblock \bibinfo{title}{Emerging coronaviruses: genome structure,
  replication, and pathogenesis}.
\newblock \bibinfo{journal}{Journal of Medical Virology} \bibinfo{volume}{92},
  \bibinfo{pages}{418--423}.
\bibitem[{Cock et~al.(2009)Cock, Antao, Chang, Chapman, Cox, Dalke, Friedberg,
  Hamelryck, Kauff, Wilczynski et~al.}]{cock2009biopython}
\bibinfo{author}{Cock, P.J.}, \bibinfo{author}{Antao, T.},
  \bibinfo{author}{Chang, J.T.}, \bibinfo{author}{Chapman, B.A.},
  \bibinfo{author}{Cox, C.J.}, \bibinfo{author}{Dalke, A.},
  \bibinfo{author}{Friedberg, I.}, \bibinfo{author}{Hamelryck, T.},
  \bibinfo{author}{Kauff, F.}, \bibinfo{author}{Wilczynski, B.}, et~al.,
  \bibinfo{year}{2009}.
\newblock \bibinfo{title}{Biopython: freely available python tools for
  computational molecular biology and bioinformatics}.
\newblock \bibinfo{journal}{Bioinformatics} \bibinfo{volume}{25},
  \bibinfo{pages}{1422--1423}.
\bibitem[{Comas et~al.(2009)Comas, Homolka, Niemann and
  Gagneux}]{comas2009genotyping}
\bibinfo{author}{Comas, I.}, \bibinfo{author}{Homolka, S.},
  \bibinfo{author}{Niemann, S.}, \bibinfo{author}{Gagneux, S.},
  \bibinfo{year}{2009}.
\newblock \bibinfo{title}{Genotyping of genetically monomorphic bacteria: {DNA}
  sequencing in mycobacterium tuberculosis highlights the limitations of
  current methodologies}.
\newblock \bibinfo{journal}{PloS One} \bibinfo{volume}{4},
  \bibinfo{pages}{e7815}.
\bibitem[{Coutard et~al.(2020)Coutard, Valle, de~Lamballerie, Canard, Seidah
  and Decroly}]{coutard2020spike}
\bibinfo{author}{Coutard, B.}, \bibinfo{author}{Valle, C.},
  \bibinfo{author}{de~Lamballerie, X.}, \bibinfo{author}{Canard, B.},
  \bibinfo{author}{Seidah, N.}, \bibinfo{author}{Decroly, E.},
  \bibinfo{year}{2020}.
\newblock \bibinfo{title}{The spike glycoprotein of the new coronavirus
  {2019-nCoV} contains a furin-like cleavage site absent in {CoV} of the same
  clade}.
\newblock \bibinfo{journal}{Antiviral Research} \bibinfo{volume}{176},
  \bibinfo{pages}{104742}.
\bibitem[{Follis et~al.(2006)Follis, York and Nunberg}]{follis2006furin}
\bibinfo{author}{Follis, K.E.}, \bibinfo{author}{York, J.},
  \bibinfo{author}{Nunberg, J.H.}, \bibinfo{year}{2006}.
\newblock \bibinfo{title}{Furin cleavage of the {SARS} coronavirus spike
  glycoprotein enhances cell--cell fusion but does not affect virion entry}.
\newblock \bibinfo{journal}{Virology} \bibinfo{volume}{350},
  \bibinfo{pages}{358--369}.
\bibitem[{Gorbalenya et~al.(2020)}]{gorbalenya2020species}
\bibinfo{author}{Gorbalenya, A.}, et~al., \bibinfo{year}{2020}.
\newblock \bibinfo{title}{The species severe acute respiratory syndrome-related
  coronavirus: classifying {2019-nCoV} and naming it {SARS-CoV-2}}.
\newblock \bibinfo{journal}{Nature Microbiology} .
\bibitem[{Guan et~al.(2020)Guan, Ni, Hu, Liang, Ou, He, Liu, Shan, Lei, Hui
  et~al.}]{guan2020clinical}
\bibinfo{author}{Guan, W.j.}, \bibinfo{author}{Ni, Z.y.}, \bibinfo{author}{Hu,
  Y.}, \bibinfo{author}{Liang, W.h.}, \bibinfo{author}{Ou, C.q.},
  \bibinfo{author}{He, J.x.}, \bibinfo{author}{Liu, L.}, \bibinfo{author}{Shan,
  H.}, \bibinfo{author}{Lei, C.l.}, \bibinfo{author}{Hui, D.S.}, et~al.,
  \bibinfo{year}{2020}.
\newblock \bibinfo{title}{Clinical characteristics of coronavirus disease 2019
  in {China}}.
\newblock \bibinfo{journal}{New England Journal of Medicine} .
\bibitem[{He et~al.(2004)He, Leeson, Ballantine, Andonov, Baker, Dobie, Li,
  Bastien, Feldmann, Strocher et~al.}]{he2004characterization}
\bibinfo{author}{He, R.}, \bibinfo{author}{Leeson, A.},
  \bibinfo{author}{Ballantine, M.}, \bibinfo{author}{Andonov, A.},
  \bibinfo{author}{Baker, L.}, \bibinfo{author}{Dobie, F.},
  \bibinfo{author}{Li, Y.}, \bibinfo{author}{Bastien, N.},
  \bibinfo{author}{Feldmann, H.}, \bibinfo{author}{Strocher, U.}, et~al.,
  \bibinfo{year}{2004}.
\newblock \bibinfo{title}{Characterization of protein--protein interactions
  between the nucleocapsid protein and membrane protein of the {SARS}
  coronavirus}.
\newblock \bibinfo{journal}{Virus Research} \bibinfo{volume}{105},
  \bibinfo{pages}{121--125}.
\bibitem[{Kim et~al.(2012)Kim, Lovell, Tiew, Mandadapu, Alliston, Battaile,
  Groutas and Chang}]{kim2012broad}
\bibinfo{author}{Kim, Y.}, \bibinfo{author}{Lovell, S.}, \bibinfo{author}{Tiew,
  K.C.}, \bibinfo{author}{Mandadapu, S.R.}, \bibinfo{author}{Alliston, K.R.},
  \bibinfo{author}{Battaile, K.P.}, \bibinfo{author}{Groutas, W.C.},
  \bibinfo{author}{Chang, K.O.}, \bibinfo{year}{2012}.
\newblock \bibinfo{title}{Broad-spectrum antivirals against {3C} or {3C}-like
  proteases of picornaviruses, noroviruses, and coronaviruses}.
\newblock \bibinfo{journal}{Journal of Virology} \bibinfo{volume}{86},
  \bibinfo{pages}{11754--11762}.
\bibitem[{Lei et~al.(2018)Lei, Kusov and Hilgenfeld}]{lei2018nsp3}
\bibinfo{author}{Lei, J.}, \bibinfo{author}{Kusov, Y.},
  \bibinfo{author}{Hilgenfeld, R.}, \bibinfo{year}{2018}.
\newblock \bibinfo{title}{Nsp3 of coronaviruses: Structures and functions of a
  large multi-domain protein}.
\newblock \bibinfo{journal}{Antiviral Research} \bibinfo{volume}{149},
  \bibinfo{pages}{58--74}.
\bibitem[{Levandowsky and Winter(1971)}]{levandowsky1971distance}
\bibinfo{author}{Levandowsky, M.}, \bibinfo{author}{Winter, D.},
  \bibinfo{year}{1971}.
\newblock \bibinfo{title}{Distance between sets}.
\newblock \bibinfo{journal}{Nature} \bibinfo{volume}{234}, \bibinfo{pages}{34}.
\bibitem[{Li et~al.(2008)Li, Wu, Yan, Ma, Wang, Zhang, Tang, Temperton, Weiss,
  Brenchley et~al.}]{li2008t}
\bibinfo{author}{Li, C.K.f.}, \bibinfo{author}{Wu, H.}, \bibinfo{author}{Yan,
  H.}, \bibinfo{author}{Ma, S.}, \bibinfo{author}{Wang, L.},
  \bibinfo{author}{Zhang, M.}, \bibinfo{author}{Tang, X.},
  \bibinfo{author}{Temperton, N.J.}, \bibinfo{author}{Weiss, R.A.},
  \bibinfo{author}{Brenchley, J.M.}, et~al., \bibinfo{year}{2008}.
\newblock \bibinfo{title}{T cell responses to whole {SARS} coronavirus in
  humans}.
\newblock \bibinfo{journal}{The Journal of Immunology} \bibinfo{volume}{181},
  \bibinfo{pages}{5490--5500}.
\bibitem[{Li(2013)}]{li2013aligning}
\bibinfo{author}{Li, H.}, \bibinfo{year}{2013}.
\newblock \bibinfo{title}{Aligning sequence reads, clone sequences and assembly
  contigs with {BWA-MEM}}.
\newblock \bibinfo{journal}{arXiv preprint arXiv:1303.3997} .
\bibitem[{Li et~al.(2005)Li, Zhang, Fu, Yu, Li, Li, Zhang, Rong, Wang, Ning
  et~al.}]{li2005sirna}
\bibinfo{author}{Li, T.}, \bibinfo{author}{Zhang, Y.}, \bibinfo{author}{Fu,
  L.}, \bibinfo{author}{Yu, C.}, \bibinfo{author}{Li, X.}, \bibinfo{author}{Li,
  Y.}, \bibinfo{author}{Zhang, X.}, \bibinfo{author}{Rong, Z.},
  \bibinfo{author}{Wang, Y.}, \bibinfo{author}{Ning, H.}, et~al.,
  \bibinfo{year}{2005}.
\newblock \bibinfo{title}{{siRNA} targeting the leader sequence of {SARS-CoV}
  inhibits virus replication}.
\newblock \bibinfo{journal}{Gene Therapy} \bibinfo{volume}{12},
  \bibinfo{pages}{751--761}.
\bibitem[{Liu et~al.(2010)Liu, Sun, Qi, Chu, Wu, Gao, Li, Yan and
  Gao}]{liu2010membrane}
\bibinfo{author}{Liu, J.}, \bibinfo{author}{Sun, Y.}, \bibinfo{author}{Qi, J.},
  \bibinfo{author}{Chu, F.}, \bibinfo{author}{Wu, H.}, \bibinfo{author}{Gao,
  F.}, \bibinfo{author}{Li, T.}, \bibinfo{author}{Yan, J.},
  \bibinfo{author}{Gao, G.F.}, \bibinfo{year}{2010}.
\newblock \bibinfo{title}{The membrane protein of severe acute respiratory
  syndrome coronavirus acts as a dominant immunogen revealed by a clustering
  region of novel functionally and structurally defined cytotoxic
  {T}-lymphocyte epitopes}.
\newblock \bibinfo{journal}{Journal of Infectious Diseases}
  \bibinfo{volume}{202}, \bibinfo{pages}{1171--1180}.
\bibitem[{Marra et~al.(2003)Marra, Jones, Astell, Holt, Brooks-Wilson,
  Butterfield, Khattra, Asano, Barber, Chan et~al.}]{marra2003genome}
\bibinfo{author}{Marra, M.A.}, \bibinfo{author}{Jones, S.J.},
  \bibinfo{author}{Astell, C.R.}, \bibinfo{author}{Holt, R.A.},
  \bibinfo{author}{Brooks-Wilson, A.}, \bibinfo{author}{Butterfield, Y.S.},
  \bibinfo{author}{Khattra, J.}, \bibinfo{author}{Asano, J.K.},
  \bibinfo{author}{Barber, S.A.}, \bibinfo{author}{Chan, S.Y.}, et~al.,
  \bibinfo{year}{2003}.
\newblock \bibinfo{title}{The genome sequence of the {SARS}-associated
  coronavirus}.
\newblock \bibinfo{journal}{Science} \bibinfo{volume}{300},
  \bibinfo{pages}{1399--1404}.
\bibitem[{McKenna et~al.(2010)McKenna, Hanna, Banks, Sivachenko, Cibulskis,
  Kernytsky, Garimella, Altshuler, Gabriel, Daly et~al.}]{mckenna2010genome}
\bibinfo{author}{McKenna, A.}, \bibinfo{author}{Hanna, M.},
  \bibinfo{author}{Banks, E.}, \bibinfo{author}{Sivachenko, A.},
  \bibinfo{author}{Cibulskis, K.}, \bibinfo{author}{Kernytsky, A.},
  \bibinfo{author}{Garimella, K.}, \bibinfo{author}{Altshuler, D.},
  \bibinfo{author}{Gabriel, S.}, \bibinfo{author}{Daly, M.}, et~al.,
  \bibinfo{year}{2010}.
\newblock \bibinfo{title}{The {Genome Analysis Toolkit}: a {MapReduce}
  framework for analyzing next-generation {DNA} sequencing data}.
\newblock \bibinfo{journal}{Genome Research} \bibinfo{volume}{20},
  \bibinfo{pages}{1297--1303}.
\bibitem[{Pfeiffer and Kirkegaard(2003)}]{pfeiffer2003single}
\bibinfo{author}{Pfeiffer, J.K.}, \bibinfo{author}{Kirkegaard, K.},
  \bibinfo{year}{2003}.
\newblock \bibinfo{title}{A single mutation in poliovirus {RNA}-dependent {RNA}
  polymerase confers resistance to mutagenic nucleotide analogs via increased
  fidelity}.
\newblock \bibinfo{journal}{Proceedings of the National Academy of Sciences}
  \bibinfo{volume}{100}, \bibinfo{pages}{7289--7294}.
\bibitem[{Prentice et~al.(2004)Prentice, McAuliffe, Lu, Subbarao and
  Denison}]{prentice2004identification}
\bibinfo{author}{Prentice, E.}, \bibinfo{author}{McAuliffe, J.},
  \bibinfo{author}{Lu, X.}, \bibinfo{author}{Subbarao, K.},
  \bibinfo{author}{Denison, M.R.}, \bibinfo{year}{2004}.
\newblock \bibinfo{title}{Identification and characterization of severe acute
  respiratory syndrome coronavirus replicase proteins}.
\newblock \bibinfo{journal}{Journal of Virology} \bibinfo{volume}{78},
  \bibinfo{pages}{9977--9986}.
\bibitem[{Ruan et~al.(2003)Ruan, Wei, Ling, Vega, Thoreau, Thoe, Chia, Ng,
  Chiu, Lim et~al.}]{ruan2003comparative}
\bibinfo{author}{Ruan, Y.}, \bibinfo{author}{Wei, C.L.}, \bibinfo{author}{Ling,
  A.E.}, \bibinfo{author}{Vega, V.B.}, \bibinfo{author}{Thoreau, H.},
  \bibinfo{author}{Thoe, S.Y.S.}, \bibinfo{author}{Chia, J.M.},
  \bibinfo{author}{Ng, P.}, \bibinfo{author}{Chiu, K.P.}, \bibinfo{author}{Lim,
  L.}, et~al., \bibinfo{year}{2003}.
\newblock \bibinfo{title}{Comparative full-length genome sequence analysis of
  14 {SARS} coronavirus isolates and common mutations associated with putative
  origins of infection}.
\newblock \bibinfo{journal}{The Lancet} \bibinfo{volume}{361},
  \bibinfo{pages}{1779--1785}.
\bibitem[{Serrano et~al.(2009)Serrano, Johnson, Chatterjee, Neuman, Joseph,
  Buchmeier, Kuhn and W{\"u}thrich}]{serrano2009nuclear}
\bibinfo{author}{Serrano, P.}, \bibinfo{author}{Johnson, M.A.},
  \bibinfo{author}{Chatterjee, A.}, \bibinfo{author}{Neuman, B.W.},
  \bibinfo{author}{Joseph, J.S.}, \bibinfo{author}{Buchmeier, M.J.},
  \bibinfo{author}{Kuhn, P.}, \bibinfo{author}{W{\"u}thrich, K.},
  \bibinfo{year}{2009}.
\newblock \bibinfo{title}{Nuclear magnetic resonance structure of the nucleic
  acid-binding domain of severe acute respiratory syndrome coronavirus
  nonstructural protein 3}.
\newblock \bibinfo{journal}{Journal of Virology} \bibinfo{volume}{83},
  \bibinfo{pages}{12998--13008}.
\bibitem[{Shu and McCauley(2017)}]{shu2017gisaid}
\bibinfo{author}{Shu, Y.}, \bibinfo{author}{McCauley, J.},
  \bibinfo{year}{2017}.
\newblock \bibinfo{title}{{GISAID}: {G}lobal initiative on sharing all
  influenza data--from vision to reality}.
\newblock \bibinfo{journal}{Eurosurveillance} \bibinfo{volume}{22}.
\bibitem[{Sievers and Higgins(2014)}]{sievers2014clustal}
\bibinfo{author}{Sievers, F.}, \bibinfo{author}{Higgins, D.G.},
  \bibinfo{year}{2014}.
\newblock \bibinfo{title}{{Clustal Omega}, accurate alignment of very large
  numbers of sequences}, in: \bibinfo{booktitle}{Multiple sequence alignment
  methods}. \bibinfo{publisher}{Springer}, pp. \bibinfo{pages}{105--116}.
\bibitem[{Sutton et~al.(2004)Sutton, Fry, Carter, Sainsbury, Walter,
  Nettleship, Berrow, Owens, Gilbert, Davidson et~al.}]{sutton2004nsp9}
\bibinfo{author}{Sutton, G.}, \bibinfo{author}{Fry, E.},
  \bibinfo{author}{Carter, L.}, \bibinfo{author}{Sainsbury, S.},
  \bibinfo{author}{Walter, T.}, \bibinfo{author}{Nettleship, J.},
  \bibinfo{author}{Berrow, N.}, \bibinfo{author}{Owens, R.},
  \bibinfo{author}{Gilbert, R.}, \bibinfo{author}{Davidson, A.}, et~al.,
  \bibinfo{year}{2004}.
\newblock \bibinfo{title}{The nsp9 replicase protein of {SARS}-coronavirus,
  structure and functional insights}.
\newblock \bibinfo{journal}{Structure} \bibinfo{volume}{12},
  \bibinfo{pages}{341--353}.
\bibitem[{Te~Velthuis et~al.(2012)Te~Velthuis, van~den Worm and
  Snijder}]{te2012sars}
\bibinfo{author}{Te~Velthuis, A.J.}, \bibinfo{author}{van~den Worm, S.H.},
  \bibinfo{author}{Snijder, E.J.}, \bibinfo{year}{2012}.
\newblock \bibinfo{title}{The {SARS}-coronavirus nsp7+ nsp8 complex is a unique
  multimeric {RNA} polymerase capable of both de novo initiation and primer
  extension}.
\newblock \bibinfo{journal}{Nucleic Acids Research} \bibinfo{volume}{40},
  \bibinfo{pages}{1737--1747}.
\bibitem[{WHO(2020)}]{who_2020}
\bibinfo{author}{WHO}, \bibinfo{year}{2020}.
\newblock \bibinfo{title}{Coronavirus disease 2019 ({COVID-19}) situation
  report – 63}.
\newblock \bibinfo{journal}{Coronavirus Disease ({COVID-2019}) Situation
  Reports} \bibinfo{volume}{00}, \bibinfo{pages}{00--00}.
\bibitem[{Wong et~al.(2004)Wong, Li, Moore, Choe and Farzan}]{wong2004193}
\bibinfo{author}{Wong, S.K.}, \bibinfo{author}{Li, W.}, \bibinfo{author}{Moore,
  M.J.}, \bibinfo{author}{Choe, H.}, \bibinfo{author}{Farzan, M.},
  \bibinfo{year}{2004}.
\newblock \bibinfo{title}{A 193-amino acid fragment of the {SARS} coronavirus
  {S} protein efficiently binds angiotensin-converting enzyme 2}.
\newblock \bibinfo{journal}{Journal of Biological Chemistry}
  \bibinfo{volume}{279}, \bibinfo{pages}{3197--3201}.
\bibitem[{Xiao et~al.(2003)Xiao, Chakraborti, Dimitrov, Gramatikoff and
  Dimitrov}]{xiao2003sars}
\bibinfo{author}{Xiao, X.}, \bibinfo{author}{Chakraborti, S.},
  \bibinfo{author}{Dimitrov, A.S.}, \bibinfo{author}{Gramatikoff, K.},
  \bibinfo{author}{Dimitrov, D.S.}, \bibinfo{year}{2003}.
\newblock \bibinfo{title}{The sars-cov s glycoprotein: expression and
  functional characterization}.
\newblock \bibinfo{journal}{Biochemical and Biophysical Research
  Communications} \bibinfo{volume}{312}, \bibinfo{pages}{1159--1164}.
\bibitem[{Yin and Yau(2019)}]{yin2019whole}
\bibinfo{author}{Yin, C.}, \bibinfo{author}{Yau, S.S.T.}, \bibinfo{year}{2019}.
\newblock \bibinfo{title}{Whole genome single nucleotide polymorphism
  genotyping of staphylococcus aureus}.
\newblock \bibinfo{journal}{Communications in Information and Systems}
  \bibinfo{volume}{19}, \bibinfo{pages}{57--80}.
\bibitem[{Yu et~al.(2017)Yu, Baune, Licinio and Wong}]{yu2017novel}
\bibinfo{author}{Yu, C.}, \bibinfo{author}{Baune, B.T.},
  \bibinfo{author}{Licinio, J.}, \bibinfo{author}{Wong, M.L.},
  \bibinfo{year}{2017}.
\newblock \bibinfo{title}{A novel strategy for clustering major depression
  individuals using whole-genome sequencing variant data}.
\newblock \bibinfo{journal}{Scientific Reports} \bibinfo{volume}{7},
  \bibinfo{pages}{44389}.
\bibitem[{Zhang et~al.(2020)Zhang, Shen, Chen and Lin}]{zhang2020origin}
\bibinfo{author}{Zhang, L.}, \bibinfo{author}{Shen, F.m.},
  \bibinfo{author}{Chen, F.}, \bibinfo{author}{Lin, Z.}, \bibinfo{year}{2020}.
\newblock \bibinfo{title}{Origin and evolution of the 2019 novel coronavirus}.
\newblock \bibinfo{journal}{Clinical Infectious Diseases} .
\bibitem[{Zhao et~al.(2005)Zhao, Cao, Zhao, Qin, Ke, Pan, Ren, Yu and
  Qi}]{zhao2005immune}
\bibinfo{author}{Zhao, P.}, \bibinfo{author}{Cao, J.}, \bibinfo{author}{Zhao,
  L.J.}, \bibinfo{author}{Qin, Z.L.}, \bibinfo{author}{Ke, J.S.},
  \bibinfo{author}{Pan, W.}, \bibinfo{author}{Ren, H.}, \bibinfo{author}{Yu,
  J.G.}, \bibinfo{author}{Qi, Z.T.}, \bibinfo{year}{2005}.
\newblock \bibinfo{title}{Immune responses against {SARS}-coronavirus
  nucleocapsid protein induced by {DNA} vaccine}.
\newblock \bibinfo{journal}{Virology} \bibinfo{volume}{331},
  \bibinfo{pages}{128--135}.

\end{thebibliography}
\end{document}